\documentclass{aa}  

\usepackage{graphicx}
\usepackage{txfonts}
\usepackage{import,pgf}
\usepackage[lofdepth,lotdepth]{subfig}

\usepackage{todonotes}
\usepackage{color}

\usepackage{soul}

\begin{document}

   \title{Deep convolutional neural networks as strong gravitational lens detectors}


   \author{C. Schaefer
          \inst{1}\thanks{christophernstrerne.schaefer@epfl.ch}
          \and
          M. Geiger\inst{1}
          \and
          T. Kuntzer\inst{1}
          \and
          J-P. Kneib\inst{1}\fnmsep\inst{2}%
          }

   \institute{ Institute of Physics, Laboratory of Astrophysics, Ecole Polytechnique Fédérale de Lausanne (EPFL), Observatoire de Sauverny,
              1250 Versoix, Switzerland \\
         \and
            Aix Marseille Université, CNRS, LAM (Laboratoire d'Astrophysique de Marseille) UMR 7326, 13388, Marseille, France
             }

   \date{Received XXX; accepted YYY}

 
  \abstract
   {Future large-scale surveys with high-resolution imaging will provide us with approximately  $10^5$ new strong galaxy-scale lenses. These strong-lensing systems will be contained in large data amounts, however, which are beyond the capacity of human experts to visually classify in an unbiased way.}
   {We present a new strong gravitational lens finder based on convolutional neural networks (CNNs). The method was applied to the strong-lensing challenge organized by the Bologna Lens Factory. It achieved first and third place, respectively, on the space-based data set and the ground-based data set. The goal was to find a fully automated lens finder for ground-based and space-based surveys that minimizes human inspection.}
   {We compared the results of our CNN architecture and three new variations ("invariant" "views" and "residual") on the simulated data of the challenge. Each method was trained separately five times on 17 000 simulated images, cross-validated using 3 000 images, and then applied to a test set with 100 000 images. We used two different metrics for evaluation, the area under the receiver operating characteristic curve (AUC) score, and the recall with no false positive ($\mathrm{Recall}_{\mathrm{0FP}}$).}
   {For ground-based data, our best method achieved an AUC score of 0.977 and a $\mathrm{Recall}_{\mathrm{0FP}}$ of 0.50. For space-based data, our best method achieved an AUC score of 0.940 and a $\mathrm{Recall}_{\mathrm{0FP}}$ of 0.32. Adding dihedral invariance to the CNN architecture diminished the overall score on space-based data, but achieved a higher no-contamination recall. We found that using committees of five CNNs produced the best recall at zero contamination and consistently scored better AUC than a single CNN.}
   {We found that for every variation of our CNN lensfinder, we achieved AUC scores close to 1 within $6\%$. A deeper network did not outperform simpler CNN models either. This indicates that more complex networks are not needed to model the simulated lenses. To verify this, more realistic lens simulations with more lens-like structures (spiral galaxies or ring galaxies) are needed to compare the performance of deeper and shallower networks.} 


   \maketitle


\section{Introduction}
  
Future strong gravitational lense (SL) studies will help further
constrain cosmology and galaxy evolution.
As of today, galaxy-scale lenses have been used successfully to constrain the Hubble constant by measuring the time-delay of lensed images of quasars independently of other measurement techniques \citep{2016Bonvin,2017Suyu}. The magnification of lensed source-objects allows observations and studies of background objects at much higher redshifts than are usually visible to telescopes \citep{2004Kneib, 2011Richard, 2015Atek}. Measurement of galaxy-scale SLs can accurately constrain the total mass of the galaxy by probing the dark matter structure. This can be used to estimate the fraction of dark matter in galaxy halos when used in combination with weak-lensing analysis \citep{2007Gavazzi} or by itself \citep{2007Jiang,2011More,2015Sonnenfeld}. It can also be used to constrain the slope of the inner mass density profile \citep{2002aTreu,2002bTreu,2008More,2009Koopmans,2016Cao} and the initial stellar mass function \citep{2010Treu,2010Ferreras,2016Leier}. One of the largest lens catalogs was produced by the Sloan Lens ACS Survey (SLACS) with about 100 observed lenses \citep{2008Bolton}. These SLs were discovered by selecting lens candidates from the spectroscopic database of the Sloan Digital Sky Survey (SDSS). Lens candidates were chosen by identifying the spectroscopic signature of two galaxies in the spectra, one galaxy at a greater distance than the other. These candidates were then verified by follow-up observation using the Hubble Space Telescope.

Historically, SLs were found serendipitously by human inspection of data. However, a systematic search by experts is too time-consuming to be a practical proposition for future large-scale surveys unless it were to involve citizen scientists. For example, the number of new lens systems from the Euclid mission \citep{2011Laureijs} and from the Large Synoptic Survey Telescope \citep{2009LSST} survey is expected to reach at least $10^5$ SLs among $10^9$ objects (LSST: \cite{2010Oguri}; HST: \cite{2014Pawase};   Euclid: \cite{2015Collett}). Similarly, the amount of SLs found by the SKA survey is expected to be on the same order of magnitude \citep{2015McKean}. 
Efficient automated gravitational lens-finding techniques are urgently needed.
   
The Spacewarps project\footnote{\url{https://spacewarps.org/}} was an attempt to use and train non-experts at lens classification. Through an interactive website, amateur scientists were trained to sort through data from CFHTLS \citep{2015Marshall}. They found 29 promising new lens candidates in the survey \citep{2016More},  but this method will likely be too slow and too much subject to human error for future data sets. Semi-automated methods like arc detectors using clustering techniques have been used with some success \citep{2004Lenzen,2007Cabanac} and have been further
improved. \citet{2014Joseph} and \citet{2016Paraficz} added machine-learning to these techniques, using a Principal Component Analysis (PCA) based approach to remove the foreground galaxy from the image and facilitate the detection of arcs. Recently, \citet{2017Petrillo} and \citet{2017Jacobs} started using convolutional neural networks (CNNs) for lens detection. CNNs belong to a class of efficient image-classifier techniques that have revolutionized image processing \citep{1998Lecun}. In astrophysics, they have been applied successfully to galaxy morphology \citep{2015Huertas-Company}, redshift estimation \citep{2016Hoyle}, and spectra analysis \citep{2014Hala}.

 The Euclid Strong-Lensing working group, in collaboration with the Bologna lens factory\footnote{\url{https://bolognalensfactory.wordpress.com/}}, has started the Galaxy-Galaxy Strong-Lensing challenge\footnote{\url{metcalf1.bo.astro.it/blf-portal/gg_challenge.html}} (GGSLC: Metcalf et al. 2017, in prep.) in light of future large-scale imaging surveys such as the Euclid
mission. The goal was to determine the best technique for finding
gravitational lenses for both ground-based and space-based imaging.

Our goal was to explore and optimize CNN architectures for lens classification. We successfully applied it to the GGSLC and were
awared first and third place in the two categories of the GGSLC. 
In this paper, we present the CNN lens finder in detail that
we created for the GGSLC challenge and discuss the advantages and disadvantages of CNN lens classifiers when applied on simulated and real data. The paper is organized as follows. 
Section~\ref{sec:theory} gives a brief overview of artificial neural networks (ANN) and CNNs and their usage in image processing. 
Section~\ref{sec:CNN} outlines the details of our algorithm implementation and the two winning CNN architectures of the challenge, while in Sect.~\ref{sec:results} we present some interesting alternative architectures.
Section~\ref{sec:conclusions} summarizes the results of the different architectures we applied to GGSLC data, and we discuss them.


\section{Theory} \label{sec:theory}
\subsection{Artificial neural network}

Artificial neural networks are machine-learning techniques inspired by the study of the human brain \citep[Hebbian learning: ][]{1950Hebb}. 
ANNs are capable of learning classification or regression tasks in $N$ dimensions by training using a set of labeled examples
. This makes them easily applicable to complex problems for which explicitly programmed solutions or mathematical models are difficult to write. The main drawback of ANNs is the computation cost of the training procedure. 
More modern training techniques coupled with advances in GPU processing power made ANNs versatile and capable of being applied to almost any data set. They are created by stacking layers of neurons together. Each neuron implements a linear combination (using weights $w_i$ and a bias $b$) of its input $\vec{x}$ followed by a nonlinear activation function $a(\vec{x})$,
\begin{equation}
    y(\vec{x}) = a\left(\sum_{i=1}^N w_i x_i+b\right),
    \label{equ:neuron}
\end{equation}
where $N$ is the dimension of the inputs.

\begin{figure}
   \centering
   \includegraphics[width=\hsize]{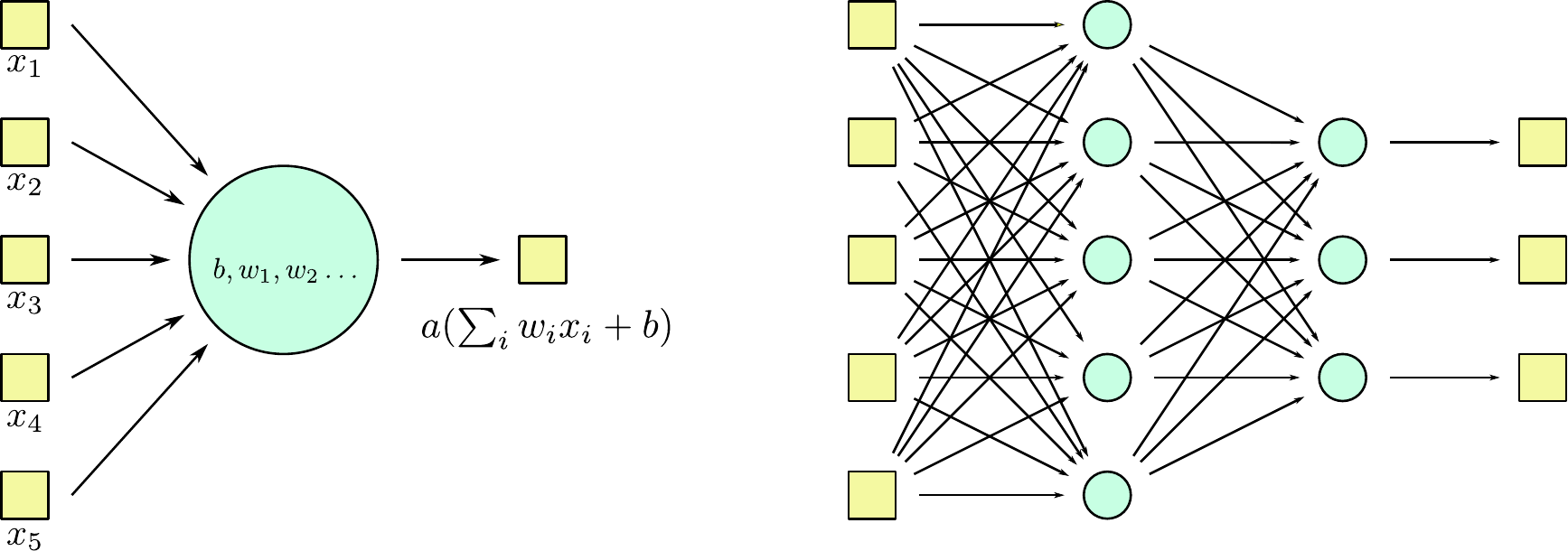}
      \caption{Left: structure of a neuronal unit. Each neuron implements a linear combination (using weights $w_i$ and a bias $b$) of its input $\vec{x}$ followed by a nonlinear activation function $a(\vec{x})$. Right: ANN structure. Neurons in the same layer all receive the same input. 
      The stacking of layers allows the ANN to define a model parametrized by the weight variables of the network.}
         \label{fig:ANN}
   \end{figure}

A layer consists of multiple neurons applied to the same input. Each output is passed as an input to the next layer.
This cascade of nonlinear combinations of inputs ends at the output layer (see Fig.~\ref{fig:ANN}).
In a classical ANN, all possible connections are established and exploited, in short, it is fully connected. A neuron in a given layer will transmit its outputs to all neurons in the next layer.
Every layer between the input and the output layer is called a hidden layer. The initial input layer is sometimes also
called a front layer. 
An ANN model is parametrized by the weights $w$ and biases $b$ of the neurons.

These weights and biases are trained iteratively. ANNs make predictions when presented with a training input.
As the model parameters are randomly initialized, the first predictions are very different from the ground truth of the input.

The ANN then evaluates the error according to some predefined cost function and computes appropriate corrections to the parameters. 
These prediction errors are propagated backward through the layers, from the output to the front layer, and induce parameter updates.
The technique is known as back-propagation \citep{1986Rumelhart} and is commonly built on gradient descent for the computation of the parameter updates.

\subsection{Convolutional neural network}

Deep ANNs, models that have more than one or two hidden layers, perform better than shallow networks. 
The mathematical evidence for this statement is still scarce, but it is empirically observed.
The continued growth in computation power made ANNs interesting for scientific application.
However, computational cost of training increases with depth, and limitations in gradient-based procedures are challenging performance obstacles.
Training with gradient methods generates a so-called vanishing-gradient problem, first identified by \citet{1991Hochreiter}.
The magnitude of the gradient diminishes as it is back-propagated through the layers.
The typical result is that layers close to the front layer effectively stop learning.
While still affected by the vanishing-gradient problems, CNNs limit its effect by reducing the number of connections and sharing weights.
This mitigation motivated the development of CNNs and their subsequent application to image recognition \citep[Lenet-5 model,][]{1998Lecun}.

The breakthrough for CNN came when \citet{2012Krizhevsky} created an architecture that won the 2012 ImageNet Large-Scale Visual Recognition Challenge\footnote{\url{http://image-net.org/}}. 
His submission achieved a classification error of only $15.3\%$  compared to the second-best submission with $26.2\%$ obtained by a method not based on a neural network. CNNs have been used extensively in image processing ever since.

  \begin{figure}
   \centering
   \includegraphics[width=\hsize]{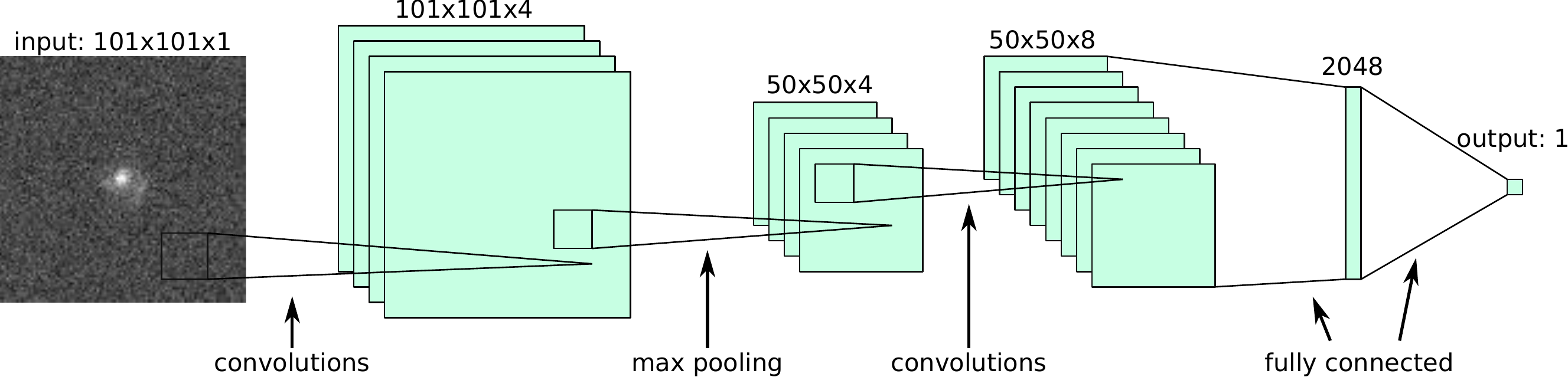}
      \caption{Example of a CNN architecture: The input image undergoes a series of convolution layers into a series of feature maps. 
      The first convolution transforms the 101x101 pixel image into four 101x101 pixel feature maps. To lower computation cost, max-pooling layers are used in between convolutions. They reduce the dimensionality of the image, dividing the size of the image by two. A fully connected layer then combines all feature maps for the classification.}
         \label{fig:CNN}
   \end{figure}


Our CNNs (Fig.~\ref{fig:CNN}) are created by stacking the following layers: convolution layers convolve the input image by a number of small kernels (or features maps, typically of dimension $3\times3$ to $7\times7$). 
The parameters to be optimized during training are the individual kernels. These weights are shared by all neurons in the layers (the kernels are the same for the whole layer).
Pooling layers reduce the dimensionality of the input to decrease the number of parameters and avoid overfitting. The most common pooling technique is the max-pooling method. It partitions the input image into non-overlapping quadrants and yields the maximum value in the quadrant. 
Fully connected (fc) layers are the classic ANN neuron layer. 
Every input is connected to every neuron of the layers. 
They are used as the final CNN layers to merge the information contained in the feature maps into the desired output form. 
Dropout layers are only active during training. They randomly sever half the connections between the two layers they separate \citep{2012Hinton}. 
This is done to reduce coadaption of the neurons (learning the same features) and reduce overfitting. 
Batch normalization layers normalize and shift the output along a small input sample $B=\{x_{1...m}\}$ following the equation 
\begin{equation}
    y_i = \gamma \dfrac{x_i -\mu_B}{\sigma^2_B} +\beta,
\end{equation}
where $\mu_B$ and $\sigma_B$ are the mean and the variance over $B$. $\gamma$ and $\beta$ are two model parameters of the layer. Batch normalization is used to increase the training speed of the CNN \citep{2015Ioffe}.

Convolution layers take advantage of the local spatial correlation in the data. 
Stacking multiple convolution layers implies a global treatment of the signal, making the network shift-invariant (i.e., features will be detected independently of their position). This make CNNs especially effective when treating images \citep{2016Mallat}.

\subsection{Data set of the Galaxy-Galaxy Strong-Lensing Challenge}

The data for the GGSLC was provided by the Bologna lens factory challenge. The Bologna lens factory project is a complex lens-simulation project.
It is based on the Millennium simulation \citep{2006Lemson}, with modeling of the gravitational lensing effect using the {\tt Glamer} ray-tracing tool \citep{2014Metcalf} and with {\tt MOKA} to create the multiplane dark halos and their substructures \citep{2012Giocoli}.
The models and the parameters used to generate the simulations were blinded for the duration of the challenge.

 Each image of the SL challenge was a 101$\times$101 pixel stamp centered around an object. 
 Participants had to submit a confidence value $p\in[0,1]$ for each image. An object with a high confidence value was interpreted as a lens.
 Two categories of data were proposed with separate data sets, each with 20\,000 training and 100\,000 test images:
(i) a space-based data set that consisted of images in a single visible band (simulating exposures of the \emph{Euclid} instrument VIS), and
(ii) a ground-based data set with images taken in four bands (U, G, R, and I) with a lower singal-to-noise ratio (S/N) and random masking of pixels, mimicking noisy data.

The ratio of lenses to non-lenses in the simulated data was much higher than in reality, around one-to-one, as an imbalance of examples (called skewed classes) can lead to biases. The results are have been made public, and a detailed discussion of the simulations and results will be provided in Metcalf et al. (in prep). 
Our baseline architecture submission to GGSLC ranked first in the space-based data category and third in the ground-based category (Fig.~ \ref{fig:RocResult}). 
CNNs in general dominated the challenge. CNN-based methods filled the seven best submission  in both categories.







\section{CNN lensfinder: architectures} \label{sec:CNN}

\begin{figure}
    \centering
    \subfloat[Baseline]{
        \centering
        \def\svgwidth{2.06cm}
        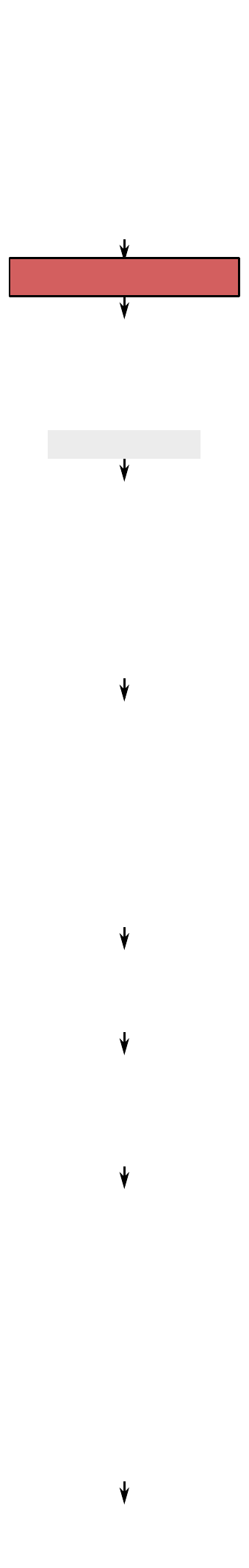
        \label{fig:Baseline}
    }\qquad\qquad
    \subfloat[Residual]{
        \centering
        \def\svgwidth{5cm}
        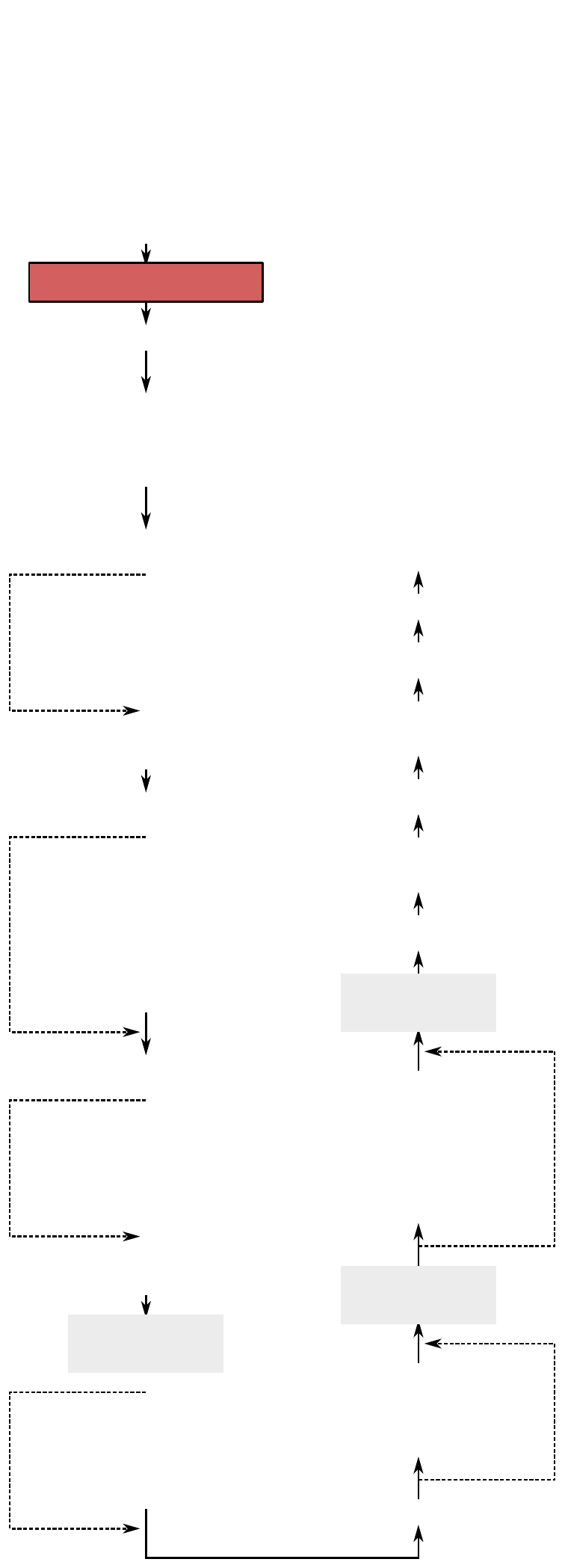
        \label{fig:Residual}
    }
    \caption{Visualization of the baseline and residual architecture for the CNN lensfinder: 
    The convolution blocks (red) indicate the size of the kernel and the number of features. 
    The fully connected blocks (yellow) indicate the number of features.     The arrows indicate the flow of the data, and between the blocks, we show the dimensionality of the input ($N_{\mathrm{pixel}}\times N_{\mathrm{pixel}} \times N_{\mathrm{features}}$). 
    The last fully connected layer yields a confidence value of the object being a lens.
    The initial layer has $N_b$ features, either one or four, depending on the category of the data (space and ground, respectively). 
    Batch normalization and dropout layers are indicated as gray blocs.}
\end{figure}

For this paper, four different types of CNN architectures were applied to the training data of the GGSLC: a simple CNN architecture that forms the baseline comparison for the paper, a so-called residual architecture based on the paper by \citet{2015He}, and two further architectures with additional invariant properties. 
The final version of each architecture was selected after a heuristic study of the parameter space. 

\subsection{Baseline architecture}

The baseline architecture as shown in Fig.~\ref{fig:Baseline} was inspired from typical CNN architectures that performed well in the ImageNet competition \citep{2014Simonyan}. It is organized by stacking convolution blocks. This simple baseline architecture achieved first place in the space component of GGSLC.
A convolution block is the superposition of 2 convolutional layers followed by a pooling layer to reduce the dimensionality of the image and a batch-normalization layer. 
The baseline architecture is comprised of 8 convolutional layers, organized into 3 convolution blocks and 2 stand-alone layers, and 3 fully connected layers at the top.
There is thus a total of 11 layers. 
With the exception of the initial layer, every convolution layer uses $3\times3$ convolution kernels for efficiency reasons \citep{2014Simonyan}. 
The first convolution layer uses a $4\times4$ kernel to yield an even number of pixels for easier manipulation. 
At each convolution block, the number of features was doubled, resulting in 256 features in the last block.
The fully connected layers used either 1024 or 2048 features.

For each layer, we chose a modified version of the rectifier linear unit (ReLU) activation function because of its sparse representation capability \citep{2011Glorot,2016Arpit}. 
The activation is given by 
\begin{equation}
    f(x) = \frac{1}{\sqrt{\pi-1}} \left( \sqrt{2 \pi} \max(0,x) - 1 \right).
    \label{equ:RELU}
\end{equation}

Inputs of the networks have dimension of $101\times101\times N_b$, where $N_b$ is the number of bands ($N_b=1$ for space and $N_b=4$ for ground).
The wavelength-dependent information (in the third dimension) is handled naturally by extending the kernel dimension from two to three (spatial to spatial plus wavelengths).

\subsection{Residual architecture}

\begin{figure}
\centering
\includegraphics[width=\hsize]{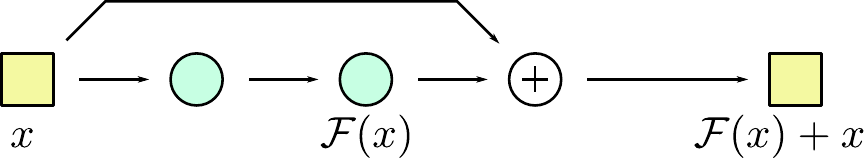}

  \caption{Structure of a residual block: The feature maps $F(\vec{x})$ from two stacked convolutional layers are added to input $\vec{x}$. Each green circle represents a convolutional layer.}
     \label{fig:Residuallayer}
\end{figure}

A common way for improving CNN is to increase the depth, that is, the number of convolutional layers. 
With creating increasingly deeper CNNs comes the vanishing-gradient problem detailed before. 
At some point in the training process, the accuracy starts to saturate and degrade, generating an upper limit to the possible depth of CNNs. 
To compensate for this, \citet{2015He} introduced residual learning. 
In the GGSLC challenge, Francois Lanusse's deep lens classifier \citep{2017Lanusse} used residual learning to create a 46-layer deep CNN that won the ground part of the challenge. 
We adapted our residual architecture to analyze the advantages and disadvantages compared to the baseline CNN. 

In a classical convolution layer, the feature map is created from scratch, that is, it learns an unreferenced mapping. 
The end-goal of the training process is to find parameters that minimize the cost function. We denote by $H(\vec{x})$ the optimum feature map and by $F(\vec{x})$ the map currently held in the parameters.
In other words, the training updates $F(\vec{x})$ until

\begin{equation}
    H(\vec{x}) = F(\vec{x}).
\end{equation}
In contrast, residual networks train by optimizing a residual mapping $\vec{x}$, or the difference between the ideal and the real feature map. 
\citet{2015He} stated that it is easier to optimize the residual feature map than the unreferenced map,
 \begin{equation}
    H(\vec{x}) = F(\vec{x}) + \vec{x},
\end{equation}
where $\vec{x}$ is the identity mapping obtained by using shortcut connections skipping the convolution layers (Fig.~\ref{fig:Residuallayer}). 
Our residual architecture as shown in Fig.~\ref{fig:Residual} is 20-layer deep with 3 small residual blocks, 4 large residual blocks, and 3 fully connected layers with 1024 features. The small residual block is composed of 2 convolutions and 1 shortcut, keeping the same number of features. The large residual block is composed of 3 convolutions and 1 shortcut followed by a convolution layer, doubling the number of features.

\subsection{Implementation details}

Other than the differences in the approach to the problem, the networks shared a number of implementation details that we outline here.

\begin{itemize}
\item Cost function: 
we chose the binary cross-entropy cost function as the cost function $C$ driving the training,
\begin{equation}
C = - \dfrac{1}{N}\sum \left\{y \ \mathrm{ln}(y_{p}) + (1-y)\left[(1- \mathrm{ln}(y_{p})\right] \right\} 
,\end{equation}
\label{equ:costentropy}
where $N$ is the number of training examples, $y$ is the ground truth, and $y_{p}$ is the classification prediction.
\item Data augmentation: 
to increase the number of training samples, we used data-augmentation techniques.
The goal is to generate more examples out of the original training set by exploiting physically invariant transformations, for example, rotating the image by 90 degrees.
The benefit of increasing the training set size is to reduce overfitting.Taking advantage of the dihedral group symmetry (Fig.~\ref{fig:Dihedral}) of the lens problem, the training sets were augmented using 90-degree rotations and flipping operations. 
We did not use rotation angles different than 90 degrees to avoid having to interpolate in pixel space.

\item Training:
the challenge training set was subdivided into a training set (of 17\,000 images) that was used by the networks to learn and a validation set (3\,000 images strong) to check the performances on an independent set.
The performance was monitored every 1000 steps by evaluating predictions made on the validation set. 
At each training step, we randomly selected batches of 30 images (15 lenses and 15 non-lenses) and ran the learning procedure for $\sim250-300$ epochs using the {\tt ADAM} minimization algorithm \citep{2015Kingma}.
We trained five networks with the same architecture and selected the best-performing individual.

\item Library:
the models were implemented using the {\tt Tensorflow} library \citep{tensorflow2015} on a GeForce GTX 1060 graphic card. The training time took approximately 1 hour/100 epochs for the baseline model and 2 hours/100 epochs for the residual model. The final prediction of the classification for the challenge on the 100\,000 test images took approximately 20 minutes.

\end{itemize}

\begin{figure}
    \centering
    \subfloat[Invariant]{
        \centering
        \def\svgwidth{2.06cm}
        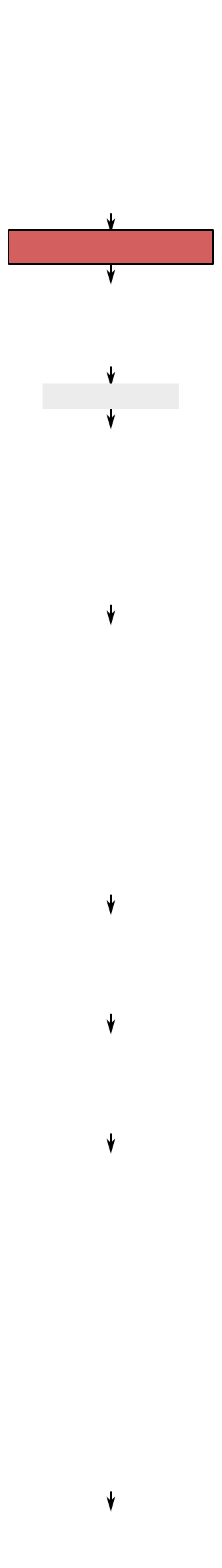
        \label{fig:Invariant}
    }\qquad\qquad
    \subfloat[Views]{
        \centering
        \def\svgwidth{5cm}
        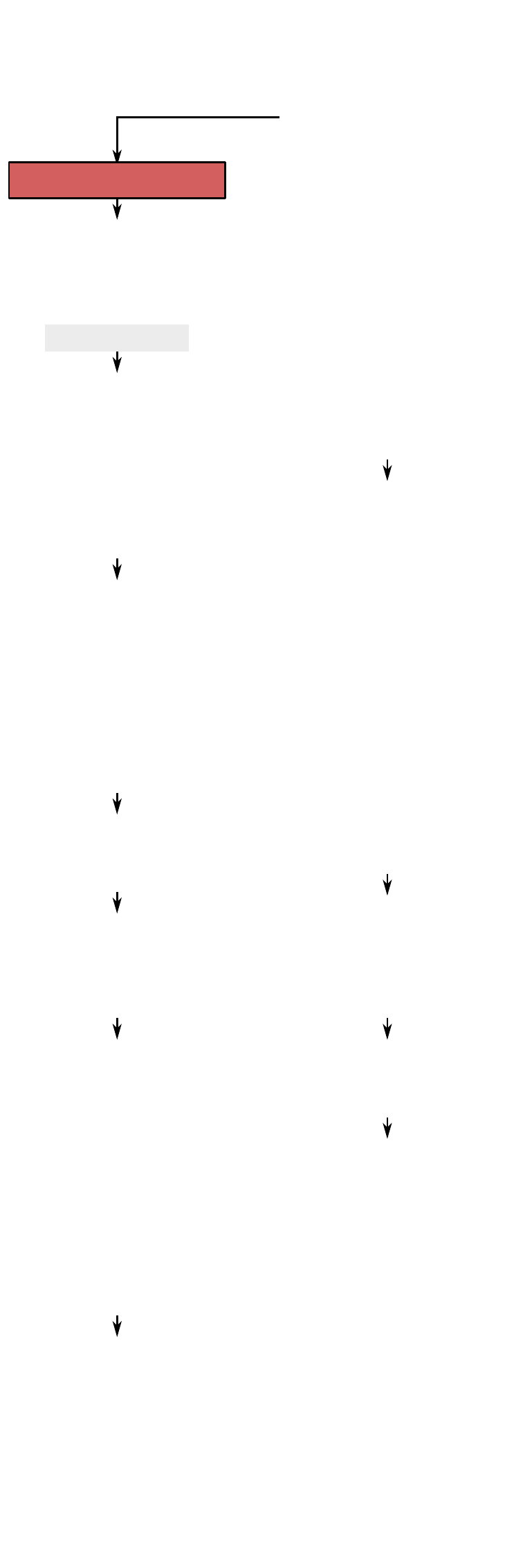
        \label{fig:Views}
    }
    \caption{Visualization of the invariant and views architecture for the CNN lensfinder: 
    The convolution blocks (red) indicate the size of the kernel and the number of features. 
    The fully connected blocks (yellow) indicate the number of features.     The arrows indicate the flow of the data, and between the blocks, we show the dimensionality of the input ($N_{\mathrm{pixel}}\times N_{\mathrm{pixel}} \times N_{\mathrm{features}}$). 
    The last fully connected layer yields a confidence value of the object being a lens.
    The initial layer has $N_b$ features, either one or four, depending on the category of the data (space and ground, respectively). 
    Batch normalization and dropout layers are indicated as gray blocs.}
\end{figure}


\subsection{Image invariance}

The idea behind these two next architectures was to deal with the inability of most lens finders to recognize and handle the invariant features of gravitational lenses. 
CNNs are, by design, already invariant to translation, but not to rotation, scaling, and flipping. The pretraining data augmentation phase renders them more robust to these symmetry operations, but not invariant. By modifying the CNN architecture so as to be invariant or more robust to different types of symmetries, we expect to reduce identification errors. The following sections describe how we increased the invariance of our models.

\subsubsection{Views architecture}

Several models trained to accomplish the same task form a committee. 
Predictions of a committee typically result in some sort of weighted combination of its members' predictions.
They have been used to improve classification results for example on the MNIST\footnote{MNIST database of handwritten digits, \url{http://yann.lecun.com/exdb/mnist/}} problem \citep{2011Ciresan} or to detect anomalies in the predictions \citep{2014Nguyen}.

The views architecture (Fig.~\ref{fig:Views}) trains two neural networks separately to look for lenses of different sizes. 
The first network looks at the whole image, detecting large lenses spanning the whole image. 
The second uses only the central part of the image.
By combining the prediction of the two networks, smaller lenses should be detected while not neglecting the detection of the larger lenses.
In other words, the first network takes as input the whole image, like the baseline model, while the second only accepts a smaller
stamp of $45\times45$ pixels.
%
To simplify the smaller network, we used only 5$\times$5 convolution layers and fewer features at each layer.


\subsubsection{Invariant architecture}

   \begin{figure}
   \centering
   \includegraphics[width=0.5\hsize]{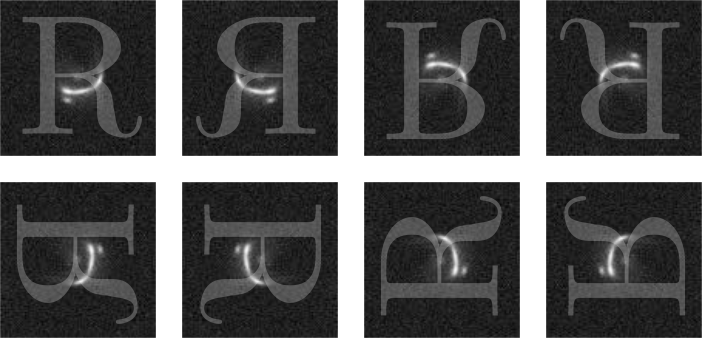}
      \caption{Representation of the dihedral symmetry group: An optimal lens finder should be invariant to the operations of this group (i.e., flipping and rotation).}
         \label{fig:Dihedral}
   \end{figure}
   
   \begin{figure}
    \centering
    \includegraphics[width=\columnwidth]{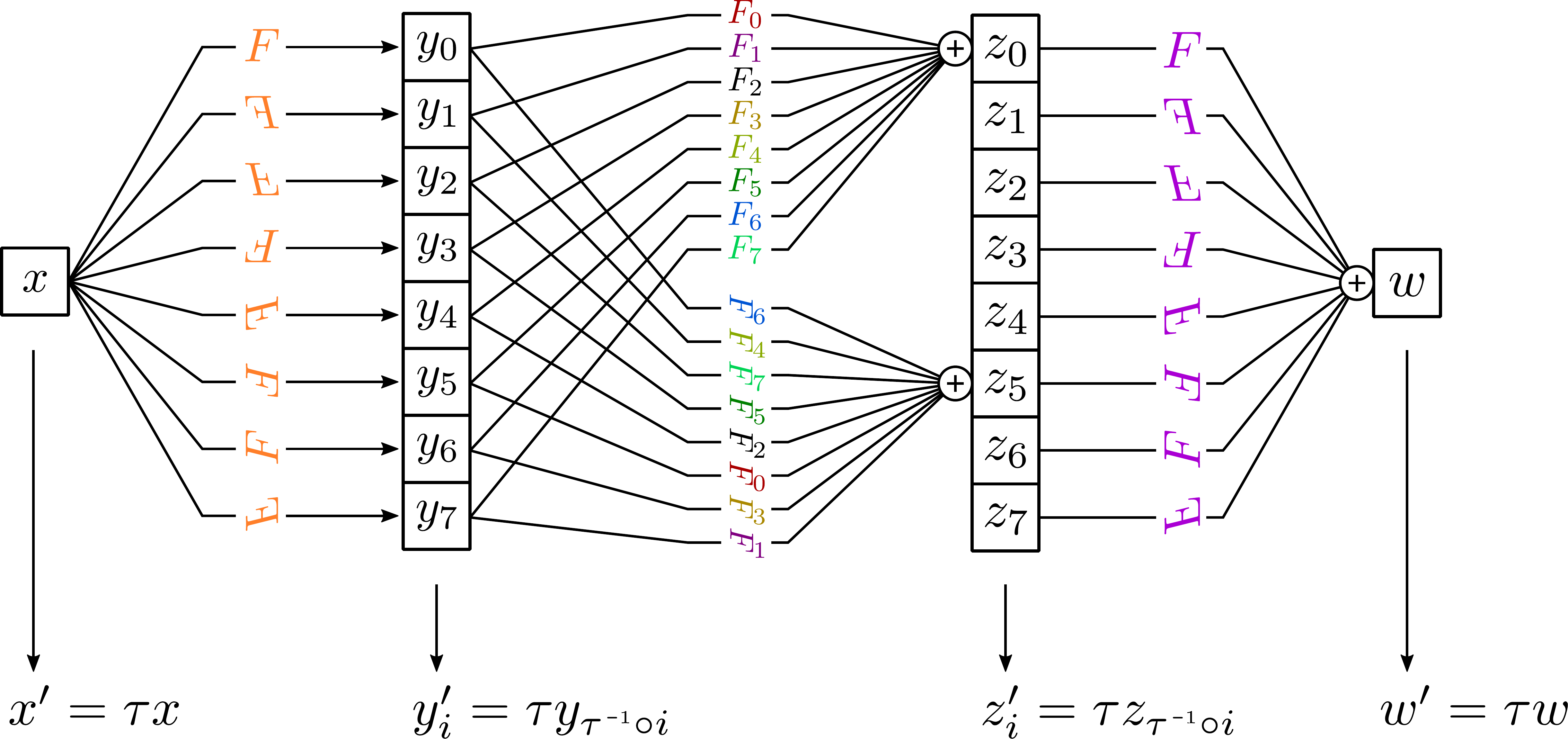}
    \caption{Dihedral equivariant architecture: Kernels with identical colors but different orientation are identical kernels to which a different dihedral operations has been applied. Phase 1: seperation into eight channels, one for every input channel and member of the dihedral group. Phase 2: convolution of the eight channels with eight separate kernels. Each output channel from a DEC layer is the sum of all the input channels convolved by all feature kernels of the layer transformed by one of the dihedral operations. Phase 3: the eight channels are summed, giving a dihedral invariant result. }
    \label{fig:dihedral_equi_layers}
\end{figure}

The invariant architecture adds additional invariant properties to the model. 
While relatively untested, this has been used with success for a galaxy morphology classifier on Galaxy Zoo data \citep{2015Dieleman,2016Dieleman}. 
The invariant architecture takes advantage of the dihedral symmetry of the lens-finding problem (Fig.~\ref{fig:Dihedral}) by using dihedral equivariant convolutional layers that we refer to as DEC layers. 

At the level of the input layer, eight operations of the same convolution kernel, transformed by a different transformation of the dihedral group, are applied to the input image. The output is divided into eight different output channels (see Fig.~\ref{fig:dihedral_equi_layers}), 
\begin{equation} \label{eq_dihedral_conv_1}
    y_i = \text{Conv}(x, F_i) \quad i \in \{0, \dots, 7\},
\end{equation}
where $i$ is one of the eight specific dihedral transformation and $M_i$ is the filter to which a dihedral transformation has been applied.

Compared to the baseline version, for the DEC layer there are eight different convolution kernel instead of one: one kernel for each transformation of the dihedral group ($F_i, \,i \in \{0,\dots, 7\}$).
Each kernel is initialized and trained separately from each other. 
Each output channel in a DEC layer is the sum of all the input channels convolved by all the different feature kernels of the layer transformed by one of the dihedral operations (Fig.~\ref{fig:dihedral_equi_layers}). 
The result of the eight channels, $y_j$, is a dihedral invariant quantity,
\begin{equation} \label{eq_dihedral_conv_2}
    y_j = \sum_{i=0}^7 \text{Conv}(x_i, j F_{j^{-1} \circ i}) \quad j \in \{0, \dots, 7\}.
\end{equation}
The two layers illustrated in Fig.~\ref{fig:dihedral_equi_layers} have the property of being invariant with respect to the dihedral group.
Our invariant architecture is shown in Fig.~ \ref{fig:Invariant} and follows the same fundamental scheme as the baseline architecture.
Since using eight channels increases computation time and makes the model more prone to overfitting, the number of features of the convolutional layers is divided by four. The invariance was tested by checking that rotated and flipped versions of the same image are attributed the same score by the classifier.







\section{Results} \label{sec:results}

In this section we describe the results of the different architectures applied to the GGSLC data. 

\subsection{Performance metric}

We first start by a brief overview of the performance metrics we used to quantify the performance of the lens classification.

\begin{figure}
\centering
    \import{Figures/}{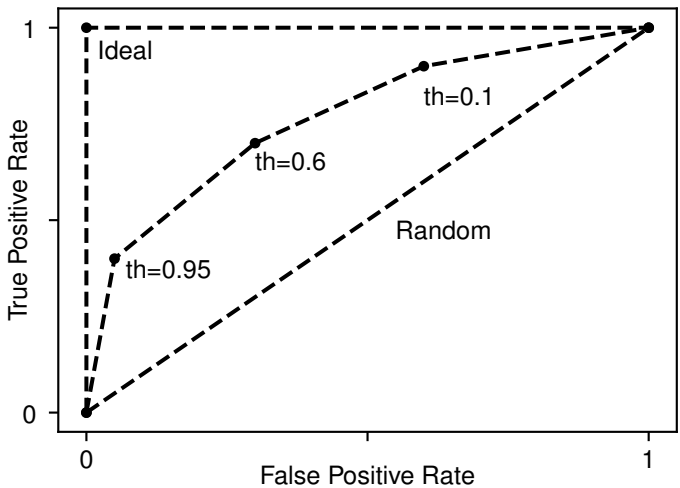}
  \caption{Receiver operating characteristic (ROC) curve: The GGSLC ranked the classifiers as a function of the area under the ROC curve (AUC). For a perfect classifier, the score is 1, and for a random classifier, it is 0.5.}     \label{fig:ROC}
\end{figure}

\begin{itemize}

\item The true-positive rate (TPR) measures how well the classifier detects lenses from the whole population of objects,
 \begin{equation}
    \mathrm{TPR} = \dfrac{N_{\mathrm{True \ positives}}}{N_{\mathrm{True \ psitives}}+N_{\mathrm{False \ negatives}}}.
\end{equation}
This metric is also known as recall. The best algorithms have a TPR close to 1.

\item The false-positive rate (FPR) measures the contamination of the positive detections by false positives,
 \begin{equation}
    \mathrm{FPR} = \dfrac{N_{\mathrm{False \ positives}}}{N_{\mathrm{True \ negatives}}+N_{\mathrm{False \ positives}}}.
\end{equation}
The best algorithms have an FPR close to 0.

\item The receiver operating characteristic (ROC) is a visual representation of the TPR and FPR. Since they depend on the threshold $ t \in (0,1) $ defined to distinguish objects as lenses or non-lenses, the ROC curve (Fig.~\ref{fig:ROC}) is created by plotting $\mathrm{TPR}(t)$ as a function of $\mathrm{FPR}(t)$ for $ t \in (0,1) $. The challenge ranked the classifiers as
a function of the area under the ROC curve (AUC), which is the integral of the ROC curve between an FPR of 0 and 1.
A perfect classifier would score 1, while a randomly predicting classifier would score 0.5.

\end{itemize}

\subsection{Training, submission, and results}

After the challenge deadline, we tested our four architectures on the GGSLC data. 
As for the baseline architecture, we used our 17\,000-image training
set and the 3000-image validation set.
Each architecture was trained five separate times. 
In Table~\ref{table:results} we show the result of this committee training. 
The performance is also evaluated in Table~\ref{table:results2} on the challenge test data as the ground-truth values were released to participants after the submission deadline.
The standard deviation of the five runs is also given. 
The two metrics used to evaluate the performance of the methods are the (i) the AUC and (ii) the zero false-positives, $\mathrm{Recall}_{\mathrm{0FP}}$ (that is, the fraction of lenses recovered with zero false-positives). 

\begin{figure}
\centering
\includegraphics[width=\hsize]{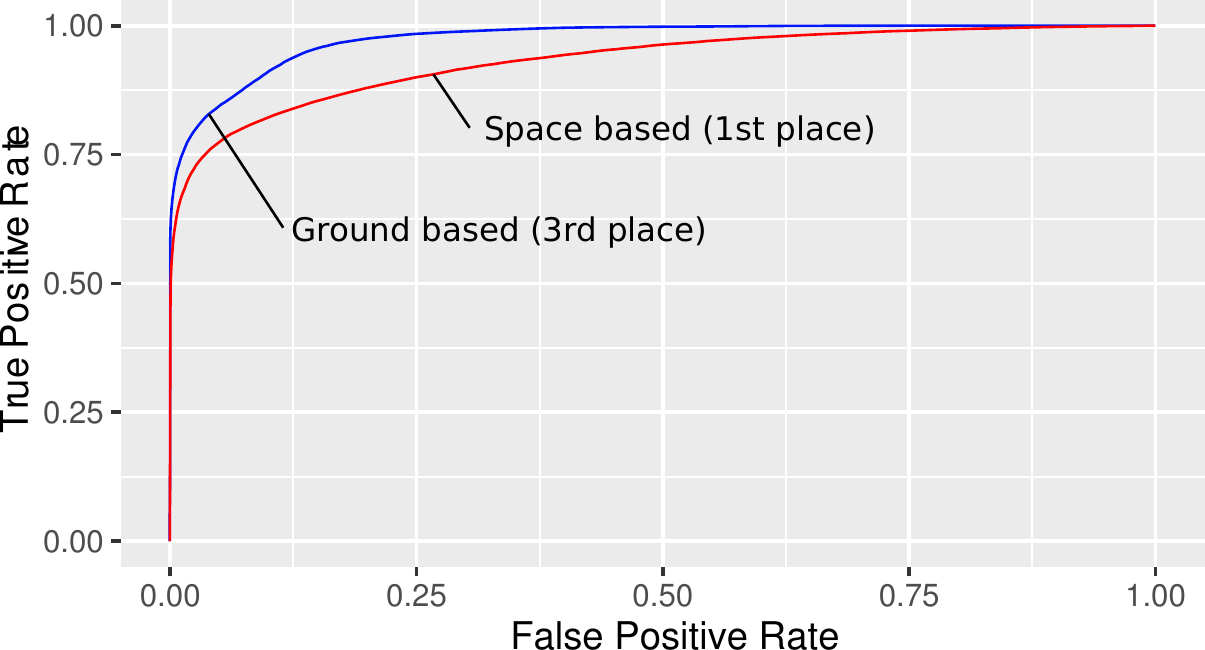}
  \caption{ROC curve of our baseline architecture submission to the GGSLC challenge. The solid line is the curve from our submission. Blue is the ground-based data category, red is the space-based data category.}
     \label{fig:RocResult}
\end{figure}

The AUC results are much better for ground-based data than for space-based data (Fig.~\ref{fig:RocResult}) although the images
have a lower S/N than the space-based images.
This increased performance could be due to the increased amount of information in the form of the four bands, instead of the single VIS-like band for space.
The lower S/N in the ground-based data does not seem to hinder prediction. 

The baseline, views, invariant and residual architectures achieved equally good AUC results on the validation set and the test set within the standard deviation of the runs. 
This is surprising because deeper networks, like the residual one, are expected to perform better than shallower models.
The scores are too close to the optimum to confidently distinguish between the architectures. 
The most likely explanation is that the simulated data were too simple for the CNN lensfinder. 
The simulations did not include spiral galaxies or some other ring-like objects capable of confusing gravitational lens classifiers. 
A more complex method was therefore not needed to classify the data correctly.
The difference that can be seen between the validation scores and the test scores in Tables \ref{table:results} and \ref{table:results2} can be attributed to a slight overfitting. This is probably due to the small size of the validation set we used in comparison to the test set.

The invariant architecture has a lower validation AUC score than the others, but performs equally well on the test set. 
This may indicate that the invariant architecture generalizes the lens model better than other architectures. 
This could be due to the imposed invariant properties, as we have given the model some additional knowledge.
This has no effect on the final test score but could become important when applying the CNN lensfinder to real data. 
Since the amount of known galaxy-scale lenses is small, a sufficiently large training set for a CNN lensfinder can only be obtained by simulated lenses \citep[see][for a CNN lensfinder applied to CHFTLS data]{2017Petrillo}. 
The caveat here is that CNNs trained on simulation might miss lenses because the simulated training set was unrealistic.
The better the CNNs generalize the lens model, the lower the
chance that they will missidentify objects. Ideas exist to force CNNs to focus on the lens model. One is to use multiple different simulations to create lenses \citep{2017Jacobs}. 
Adding dihedral invariance to CNNs could be another way of doing this.




\begin{table}
\caption{Training results: Each architecture was run five separate times. The training and validation AUC scores are the mean of these runs. The error is the standard deviation.}             
\label{table:results}      
\centering                          
\begin{tabular}{c c c}        
\hline\hline                 
Space & Training & Validation \\    
\hline                        
   baseline & $0.9920\pm0.0020$ & $0.9764\pm0.0017$ \\      
   views &  $0.9898\pm0.0030$ &  $0.9753\pm0.0021$   \\
   residual & $0.9958\pm0.0028$ & $0.9765\pm0.0024$      \\
   invariant & $0.9997\pm0.0003$ & $0.9719\pm0.0016$   \\
\hline    
Ground & Training & Validation \\    
\hline                        
   baseline & $0.9953\pm0.0090$ & $0.9905\pm0.0082$   \\      
   views & $0.9980\pm0.0010$ & $0.9924\pm0.0006$\\
   residual & $0.9990\pm0.0023$ & $0.9932\pm0.0027$  \\
   invariant & $0.9999\pm3\times 10^{-6}$ & $0.9880\pm0.0030$  \\
\hline       
\end{tabular}
\end{table}

\begin{table}
\caption{Confusion matrix (baseline architecture, GGSLC challenge) for TPR0. The TPR0 threshold was chosen by the GGSLC organizers for no false-positive in the first 10\,000 images of the test set.)}             
\label{table:confumatrix}      
\centering                          
\begin{tabular}{c c c}        
\hline\hline                 
 Space-based & Classified as non-lens & Classified as lens \\    
\hline                        
   Non-lens & 59742 & 40  \\      
   lens & 20957 & 19264     \\
\hline\hline                 
Ground-based & Classified as non-lens & Classified as lens \\    
\hline                        
   Non-lens & 50042 & 17  \\      
   lens & 21754 & 28194     \\
\hline    
\end{tabular}
\end{table}

\begin{table}
\caption{Test, $\mathrm{Recall}_{\mathrm{0FP}}$ , and $\mathrm{Recall}_{\mathrm{1FP}}$ results. Each architecture was run five times. The test scores are the mean of these runs.}             
\label{table:results2}      
\centering                          
\begin{tabular}{c c c c}        
\hline\hline                 
Space & Test AUC & $\mathrm{Recall}_{\mathrm{0FP}}$ & $\mathrm{Recall}_{\mathrm{1FP}}$\\    
\hline                        
   baseline & $0.9322\pm0.0016$ & $0.01\pm0.02$ & $0.04\pm0.04$\\      
   committee b. & $0.9326$ & $0.01$ & $0.01$  \\
   \hline  
   views &  $0.9324\pm0.0013$ &  $0.26\pm0.06$   & $0.28\pm0.07$\\
   committee v. & $0.9343$ & $0.30$ & $0.32$  \\
   \hline  
   residual & $0.9322\pm0.0006$ & $0.23\pm0.04$   & $0.29\pm0.03$   \\
   committee r. & $0.9346$ & $0.29$ & $0.30$  \\
   \hline  
   invariant & $ 0.9332\pm0.0006 $ & $0.27\pm0.04$   & $0.28\pm0.05$\\
   committee i. & $ 0.9399$ & $ 0.32 $  & $0.33$ \\
   \hline  
\hline    
Ground & Test AUC & $\mathrm{Recall}_{\mathrm{0FP}}$ & $\mathrm{Recall}_{\mathrm{1FP}}$ \\    
\hline                        
   baseline & $0.9761\pm0.0011$ & $0.44\pm0.13$ & $0.49\pm0.08$\\      
   committee b. & $0.9773$ & $0.50$ & $ 0.55 $ \\
   \hline  
   views &  $0.9746\pm0.0011$ &  $0.35\pm0.19$  & $0.43\pm0.17$ \\
   committee v. & $0.9759$ & $0.35$ & $0.39$ \\
   \hline  
   residual & $0.9775\pm0.0006$ & $0.44\pm0.06$ & $0.46\pm0.07$ \\
   committee r. & $0.9795$ & $0.50$ & $ 0.55 $ \\
   \hline  
   invariant & $0.9774\pm0.002$ & $0.39\pm0.11$ & $0.45\pm0.05$ \\
   committee i. & $0.9813 $ & $0.49$  & $0.49$ \\
   \hline  
\end{tabular}
\end{table}



\begin{figure}
    \centering
    \import{Figures/}{roc_ground.pgf}
    \caption{Logaritmic ROC curves on ground-based data. Training (dotted line), validation (half-dotted line) and test (solid line) score of all four architectures. Data come from the best of five runs in terms of validation set score.}
    \label{fig:roc_baseline_ground}
\end{figure}

\begin{figure}
    \centering
    \import{Figures/}{roc_space.pgf}
    \caption{Logaritmic ROC curves on space-based data. Training (dotted line), validation (half-dotted line) and test (solid line) score of all four architectures. Data come from the best of five runs in terms of validation set score.}
    \label{fig:roc_baseline_space}
\end{figure}

\begin{figure}
    \centering
    \import{Figures/}{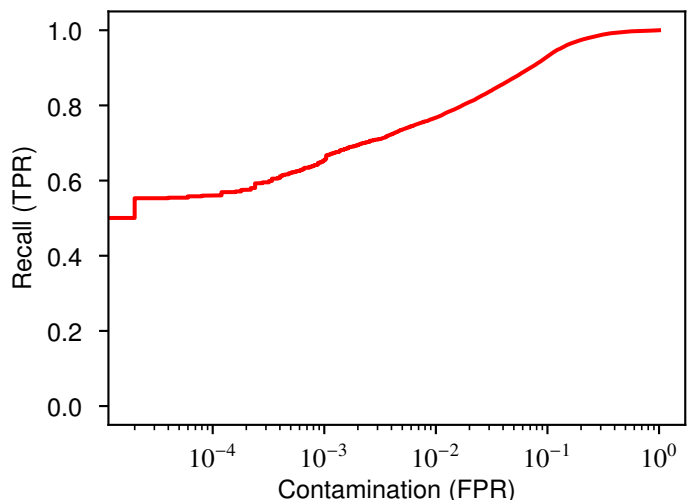}
    \caption{Logaritmic ROC curve of the baseline committee on ground-based data. The curve is the result of the baseline committee (five baseline CNNs taken together). The shaded areas represent the minimum and maximum values from the five stand-alone baseline CNNs.}
    \label{fig:ground_commi}
\end{figure}
\begin{figure}
    \centering
    \import{Figures/}{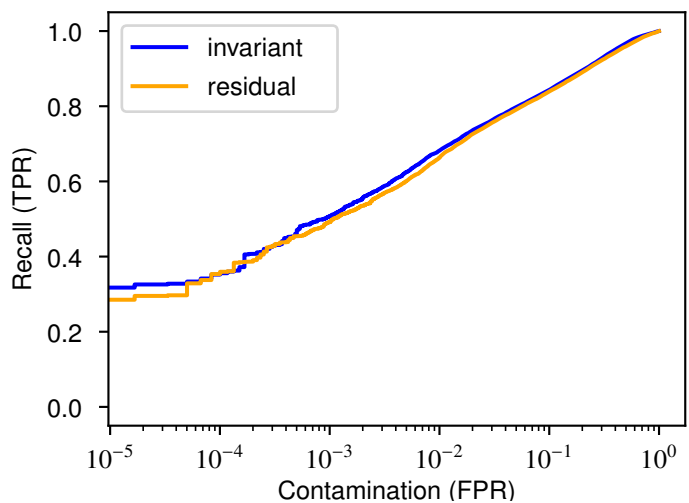}
    \caption{Logaritmic ROC curve of the invariant and residual committee on space-based data. The curve is the result of the committee (five invariant or residual CNNs taken together). The shaded areas represent the minimum and maximum values from the five stand-alone invariant or residual CNNs.}
    \label{fig:space_commi}
\end{figure}


The ground-based results are extremely encouraging, especially because of the purity of the score. 
In a classification problem with a 1-to-1000 ratio between lenses and non-lenses, 
algorithms with even a very small contamination can be dominated by false positives.
The baseline model performs well on the metric $\mathrm{Recall}_{\mathrm{0FP}}=0.44$ in the ground-based test set (Table~\ref{table:results2}).
In a more realistic setting with a ratio of lenses to non-lens objects, we would have found 22 out of the 50 lenses in a test set containing 100\,000 images without any false positives.


Table~\ref{table:results2} shows that the standard deviation of $\mathrm{Recall}_{\mathrm{0FP}}$ is large. 
Using the CNN that performed best on the validation set does not guarantee the best $\mathrm{Recall}_{\mathrm{0FP}}$ or even best AUC score (Figs.~\ref{fig:roc_baseline_ground} and~\ref{fig:roc_baseline_space}). 
The metrics vary depending on the individual result of the training run. To mitigate this, we grouped the five training runs of our model in a committee of CNNs. 
The committee output is taken as the average of their prediction. 
By compensating for each other's shortcomings, committees stabilize the results and achieve a better-than-average result for the AUC metric as well as for the $\mathrm{Recall}_{\mathrm{0FP}}$ metric (Table~\ref{table:results2}, Figs.~\ref{fig:ground_commi} and~\ref{fig:space_commi}). 
The invariant model especially is improved by this and obtains the best scores for the space-based data.

\section{Conclusions} \label{sec:conclusions}
We presented a strong gravitational lens finder based on convolutional neural networks (CNN). The method showed strong performances on simulated data. 
It won the first place and third place in the Strong Gravitational Lens Challenge (GGSLC), respectively, in the space-based and ground-based data category. We have also presented three other variations of that lensfinder, among which, a residual CNN based on the recent architecture developed by \citet{2015He}.

We found that CNNs perform better on ground-based data than on space-based data despite the lower S/N. 
This is probably due to the additional bands, which add information, but this still has to be confirmed. 
This can be done, for instance, by limiting the ground-based data to one band and comparing to the other results. 
All four CNNs achieved almost perfect ROC curve scores on the simulated data, with the highest area under the ROC curve (AUC) score up to 0.9775 for ground-based and 0.933 for space-based data. 
They also achieved a recall with a zero false-positive ($\mathrm{Recall}_{\mathrm{0FP}}$) of $50\%$ for ground-based and of $32\%$ for space-based data. 
We showed that the best $\mathrm{Recall}_{\mathrm{0FP}}$ results were achieved by committees of CNNs instead of single CNNs. Committees of CNNs consistently scored the best AUC scores. 
We also observed that adding rotation invariance to CNNs grouped together in committees produces the best space-based $\mathrm{Recall}_{\mathrm{0FP}}$ score.

Because all results are almost equally good, more conclusions about the best CNN model cannot be drawn. 
Most likely the simulations did not include enough lens-like objects capable of inducing false positives in the lensfinder, that is, the simulations were likely not realistic enough. This might explain why, contrary to expectations, the residual CNN has not performed better than the others. We will further explore CNN algorithms in the future GGSLC.

\begin{acknowledgements}
      The authors would like to acknowledge Frederic Courbin, Colin Jacobs, Ben Metcalf, and Markus Rexroth for their help and advice on the subject. CS and JPK acknowledge support from the ERC advanced grant LIDA. TK acknowledges support from the Swiss National Science Foundation.
\end{acknowledgements}


\bibliographystyle{printer.bst}
\bibliography{report.bib}

\end{document}